\documentclass[12pt]{article}

\usepackage{graphicx}
\usepackage{color}
\usepackage[colorlinks=true, urlcolor=blue, linkcolor=blue, 
citecolor=blue]{hyperref}
\usepackage{cite}

\newcommand{\beq}{\begin{equation}}
\newcommand{\eeq}{\end{equation}}
\newcommand{\beqa}{\begin{eqnarray}}
\newcommand{\eeqa}{\end{eqnarray}}
\newcommand{\beqar}{\begin{eqnarray*}}
\newcommand{\eeqar}{\end{eqnarray*}}

\begin{tiny}

\end{tiny} 

\begin{document}
\thispagestyle{empty}
$\,$

\vspace{32pt}

\begin{center}

\textbf{\Large The solar disk at high energies}

\vspace{50pt}
Miguel Guti\'errez, Manuel Masip, Sergio Mu\~noz 
\vspace{16pt}

\textit{CAFPE and Departamento de F{\'\i}sica Te\'orica y del Cosmos}\\
\textit{Universidad de Granada, E-18071 Granada, Spain}\\
\vspace{16pt}

\texttt{mgg,masip@ugr.es\hspace{0.5cm} sergiomuni07@correo.ugr.es}

\end{center}

\vspace{30pt}

\date{\today}

\begin{abstract}

High energy cosmic rays {\it illuminate} the Sun and produce an image that could be observed in up to five different channels:
a cosmic ray shadow (whose energy dependence has been studied by HAWC);
a gamma ray flux (observed at $E\le 200$ GeV by Fermi-LAT);
a muon shadow (detected by ANTARES and IceCube);
a neutron flux (undetected, as there are no hadronic calorimeters in space); and
a flux of high energy neutrinos. Since these signals are correlated, the ones already observed can be 
used to reduce the uncertainty in the still undetected ones. Here we define a simple set up that 
explains the Fermi-LAT and HAWC observations and implies very definite fluxes of neutrons and neutrinos from 
the solar disk. In particular, we provide a fit of the neutrino flux at 10 GeV--10 TeV that includes its
dependence on the zenith angle and on the period of the solar cycle.  This flux represents a 
{\it neutrino floor} in indirect dark matter searches. We show that in some benchmark models the
current bounds on the dark matter-nucleon
cross section push the solar signal below this neutrino floor.

\end{abstract}

\vfill
\eject

\section{Introduction} 
The surface of the Sun is at a temperature $T\approx 0.5$ eV, while its core is 
burning Hydrogen at $T\approx 1$ keV. Nuclear reactions there produce neutrinos that 
reach the Earth unscattered with energies of up to 10 MeV. In addition, 
solar flares are able to accelerate nuclei and electrons up to a couple of GeV. The Sun, however, can also
be observed at energies above the GeV. The emission in these other channels is indirect: instead of
particles accelerated by the Sun, it appears when high energy cosmic rays (CRs) {\it illuminate} its surface.
In particular, EGRET \cite{Orlando:2008uk} and Fermi-LAT \cite{Abdo:2011xn} 
(see also \cite{Linden:2018exo}) have observed a sustained flux of gamma rays coming 
from the solar disk that extends up to 200 GeV. 
The signal, stronger during a solar minimum and interpreted as the albedo flux produced by CRs 
showering in the Sun's surface,
is ten times above the diffuse gamma-ray background and
six times larger than a 1991 estimate by Seckel, Stanev and Gaisser \cite{Seckel:1991ffa}.
Obviously, the same mechanism should produce as well neutrinos and neutrons, which are also
neutral and thus able to reach the Earth revealing their source. 

Although the solar emission of high energy particles induced by CRs was already discussed 40 years ago\cite{Seckel:1991ffa}, 
a precise calculation has been plagued by the uncertainties introduced by the solar magnetism. Here we propose a 
complete and consistent framework that avoids these difficulties by using the data and implies a clear
correlation among the different 
signals that may be accessible at several astroparticle observatories. If observed, they would provide
a multi-messenger picture of the Sun complementary to the one obtained with light and keV-MeV 
neutrinos.

\section{Absorption of cosmic rays} 
If we point with a detector of CRs to the Sun, we will observe a shadow: the CR shadow of the Sun \cite{TibetASgamma:2013rjm}. 
Suppose there were no
solar magnetism, so that CRs follow straight lines. Then the trajectories aiming to the Earth but 
absorbed by the Sun would define a black disk of radius $r=0.26^\circ$, the angular size of the Sun as seen from 
the Earth. Indded, this is what we will see
at very high energies, when the deflection of CRs by the solar magnetic field is negligible, but not at lower energies.
CRs of energy below 100 TeV are very affected by a magnetic field that, unfortunately, is
very involved. 
First of all, it has a radial component (open lines that define the Parker interplanetary field \cite{Tautz:2010vk}) 
that grows like $1/R^2$ as we approach the surface. This gradient in the field may induce a magnetic mirror effect: CRs
approaching the Sun tend to bounce back. In addition, the solar wind induces convection, {\it i.e.}, CRs are propagating in a
plasma that moves away from the Sun and makes it more difficult to reach the surface. Finally, closer to the Sun the magnetic
turbulence increases and there appears a new type of field lines that start and end on the solar surface.
Hopefully, we can understand the absorption rate of CRs by the Sun with no need to solve these details,
just by using the data on its CR shadow together with Liouville's theorem.

The data is provided by HAWC \cite{Enriquez:2015nva}, that has studied the energy-dependence of the CR shadow during
a solar maximum. The shadow appears
at 2 TeV; it is not a black disk of $r=0.26^\circ$ but a deficit that extends into a larger angular region. By integrating it we 
find that at 2 TeV it accounts for a 6\% of a black  disk, the deficit grows to a 27\% at 8 TeV, and at 50 TeV it becomes 
a 100\% deficit, {\it i.e.}, a complete solar black disk diluted in a $2^\circ$ circle.

\begin{figure}[!t]
\begin{center}
\includegraphics[scale=0.4]{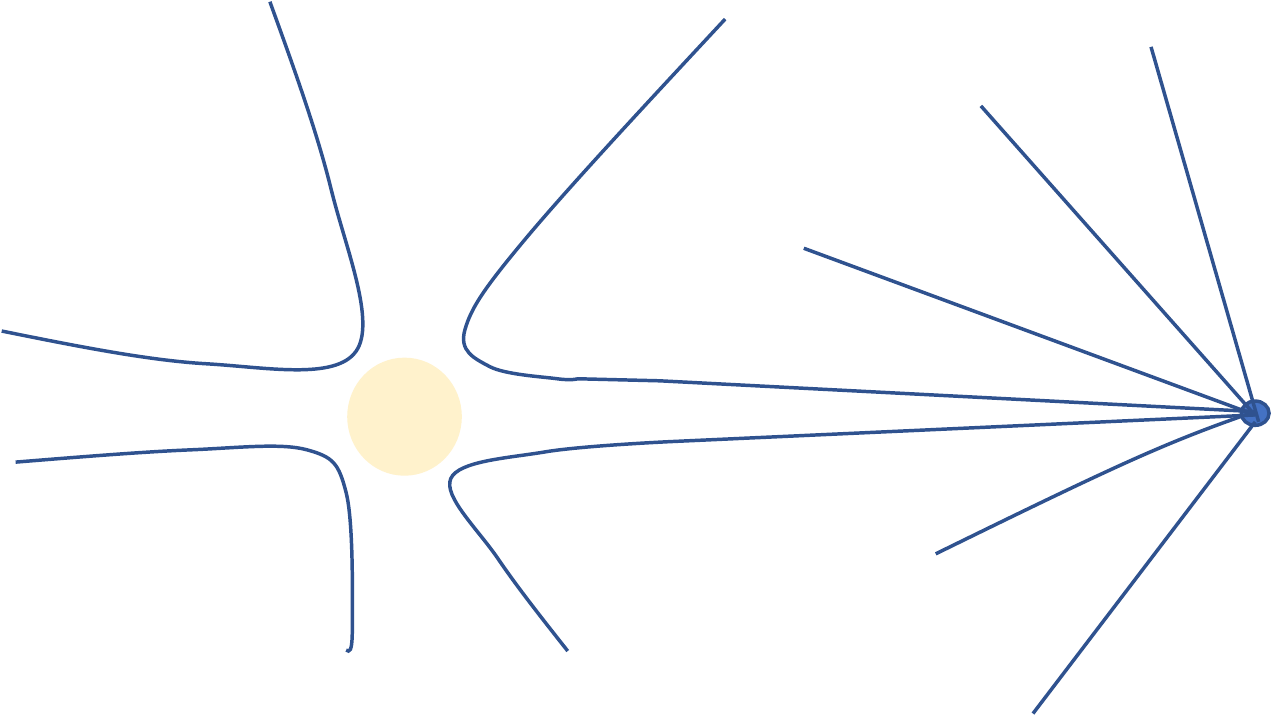}\hspace{2cm}
\includegraphics[scale=0.4]{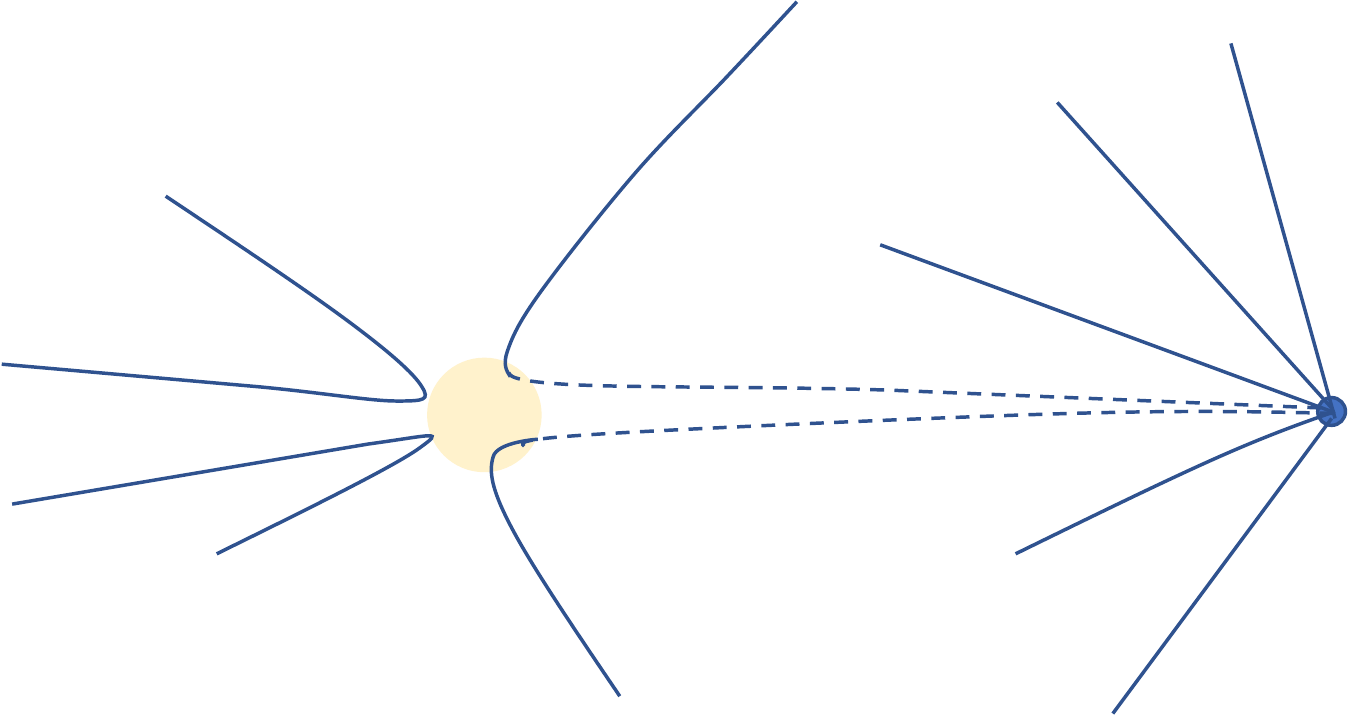}
\end{center}
\vspace{-0.3cm}
\caption{Schematic CR trajectories in the vicinity of the Sun. The solar magnetism does not create anisotropies in the
flux reaching the Earth,
but the average depth of solar matter crossed by CRs grows with the energy, increasing 
their probability to be absorbed and thus the integrated CR shadow observed by HAWC.
\label{f1}}
\end{figure}

HAWCs data suggest a simple interpretation based on Liouville's theorem. 
The theorem implies that an isotropic CR flux crossing the solar
magnetic field will stay isotropic, and that the only possible effect of the Sun is to interrupt some of the 
trajectories that were aiming to the Earth. As we illustrate in Fig.~\ref{f1}, the solar magnetic field
deflects some of the trajectories that were directed to the Earth, but other trajectories will now reach us and the net
effect should be zero: an isotropic flux crossing a static magnetic lens, including a mirror, will stay isotropic, and the only
possible effect is to create a shadow. 
At low energies HAWC sees no shadow, meaning that a negligible fraction of the CR flux reaches the solar surface. 
At higher energies, however, 
CRs that were supposed to reach the detector hit before the Sun and are absorbed (Fig.~\ref{f1}-right). 
Therefore, studying the shadow we may deduce 
the average depth of solar matter crossed by CR's of different energy in their way to the Earth.

If a CR proton crosses an average depth of $\Delta X_{\rm H}(E)$ 
the probability to be absorbed is
\beq
p^{\rm H}_{\rm abs}=1-\exp{\left(-{\Delta X_{\rm H}\over \lambda_{\rm int}^{\rm H}}\right)}\,.
\eeq
To explain the data we will assume 
\beq
{\Delta X_{\rm H}\over \lambda_{\rm int}^{\rm H}}=b_H \,E^{1.11}\,,
\eeq
with $E$ in GeV and a time dependent parameter $b_H$ that oscillates between $1.6\times 10^{-5}$ 
during a solar maximum 
and $4.8\times 10^{-5}$ during a minimum. Since the trajectory of a CR only depends on its rigidity, He nuclei 
of twice the energy cross the same average depth and
\beq
b_{\rm He} = {b_{\rm H}\over 2^{1.1}} \,{\sigma_{{\rm He}}(E)\over  \sigma_{H}(E/2)}\,.
\eeq
\begin{figure}[!t]
\begin{center}
\includegraphics[scale=0.58]{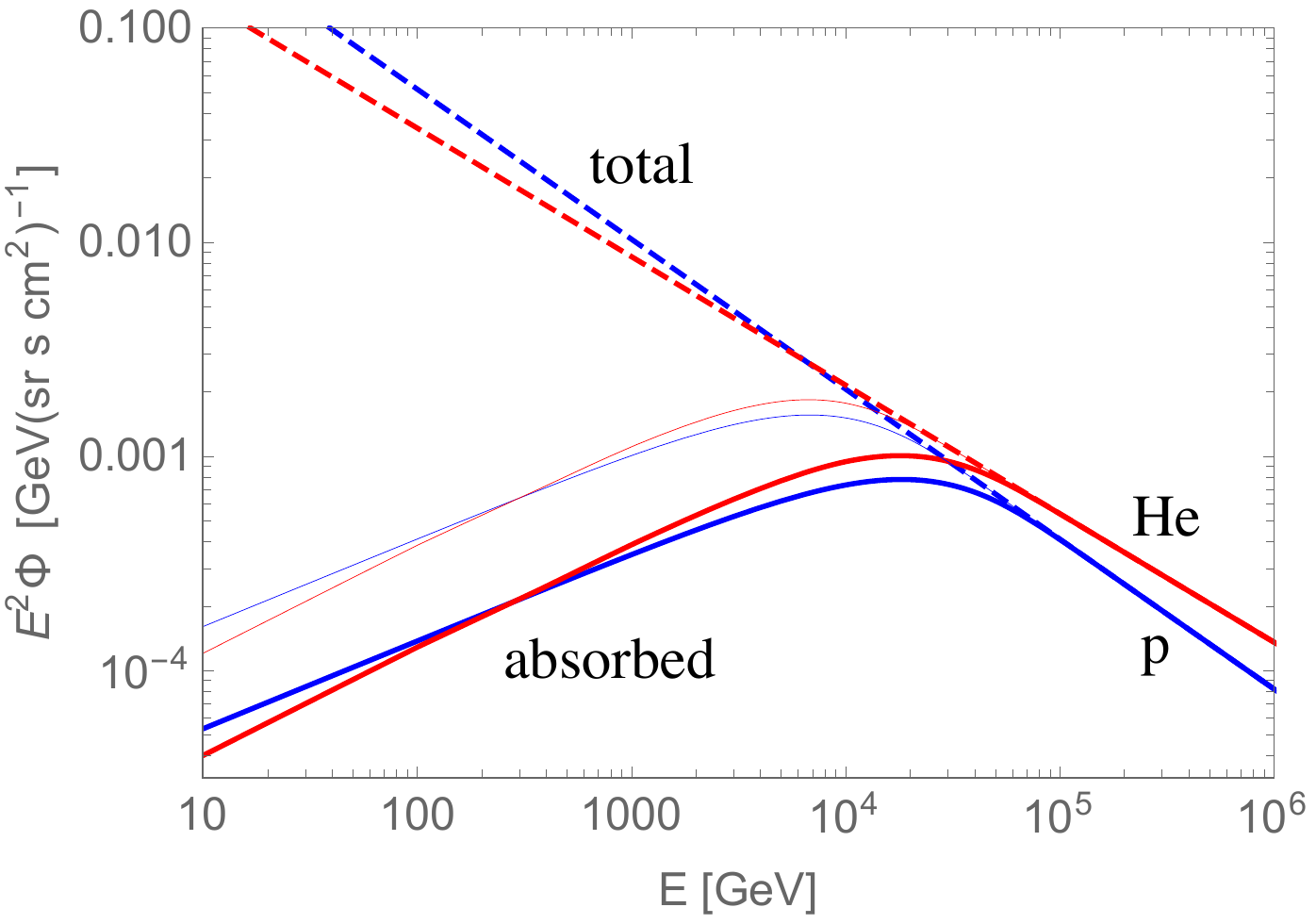}\hspace{0.7cm} \includegraphics[scale=0.65]{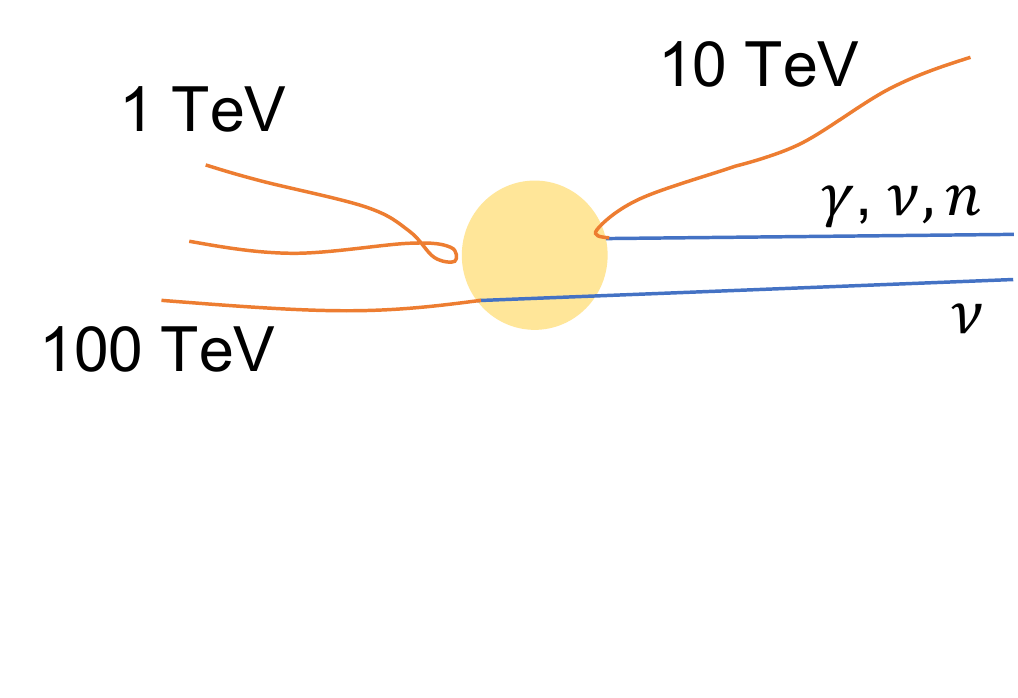}
\end{center}
\vspace{-0.3cm}
\caption{Absorbed proton and He fluxes during a solar maximum (thick) and a solar minimum (thin). On the right, 
typical CR trajectories at different energies.
\label{f2}}
\end{figure}
In Fig.~\ref{f2} we plot the CR fluxes from the solar disk that are absorbed at different
energies. We have considered a primary CR flux with only proton and He nuclei and slightly different spectral indexs
($-2.7$ and $-2.6$, respectively). 

Next we need to model the showering of these absorbed fluxes. A numerical simulation 
shows that at TeV energies only trajectories very aligned with the open field lines are able to reach the Sun's
surface. Once there, CRs will shower; some of the secondaries will be emitted 
inwards, towards the Sun, but others will be emitted outwards and may eventually reach the
Earth. The probability that a secondary particle contributes to the solar albedo flux will depend on how deep it is 
produced and in which direction it is emitted.

We assume that secondaries produced by a parent of energy $E$ above some critical 
energy $E_{\rm c}$ 
that varies between
6 TeV and 3 TeV for an active or quiet Sun, respectively, will most likely be emitted towards the Sun, whereas 
lower energy primaries will exit in a random direction:\footnote{Notice that $E_{\rm c}$ is
a factor of 2 larger when the parent particle is a He nucleus.}
\beq
p_{\rm out}={1\over 2}\, \exp \left( {E\over E_{\rm c}} \right)^2.
\eeq
Accordingly, we also assume that charged particles of energy below $E_c$ are unable to keep penetrating the Sun:
they are trapped by closed magnetic lines at the depth where they are produced and shower horizontally.

Under these hypothesis, we use cascade equations \cite{Gaisser:1990vg}
to find the final albedo flux of neutral particles (see the complete equations in \cite{Masip:2017gvw,Gamez:2019dex}). 
The key difference with the usual showers in the Earth's atmosphere is due to the thin environment where these
solar showers develop: below the optical surface of the Sun, it takes 1500 km to cross 
just 100 g/cm$^2$. As a consequence, TeV pions and even muons decay before they loose energy, defining 
photon and neutrino fluxes well above the atmospheric ones. For neutrinos, to the albedo flux we must add the 
neutrinos produced in the opposite side of the Sun \cite{Edsjo:2017kjk,Ng:2017aur,Arguelles:2017eao}.
Our results for the signal in the different channels are the following.

\begin{figure}[!t]
\begin{center}
\includegraphics[scale=0.53]{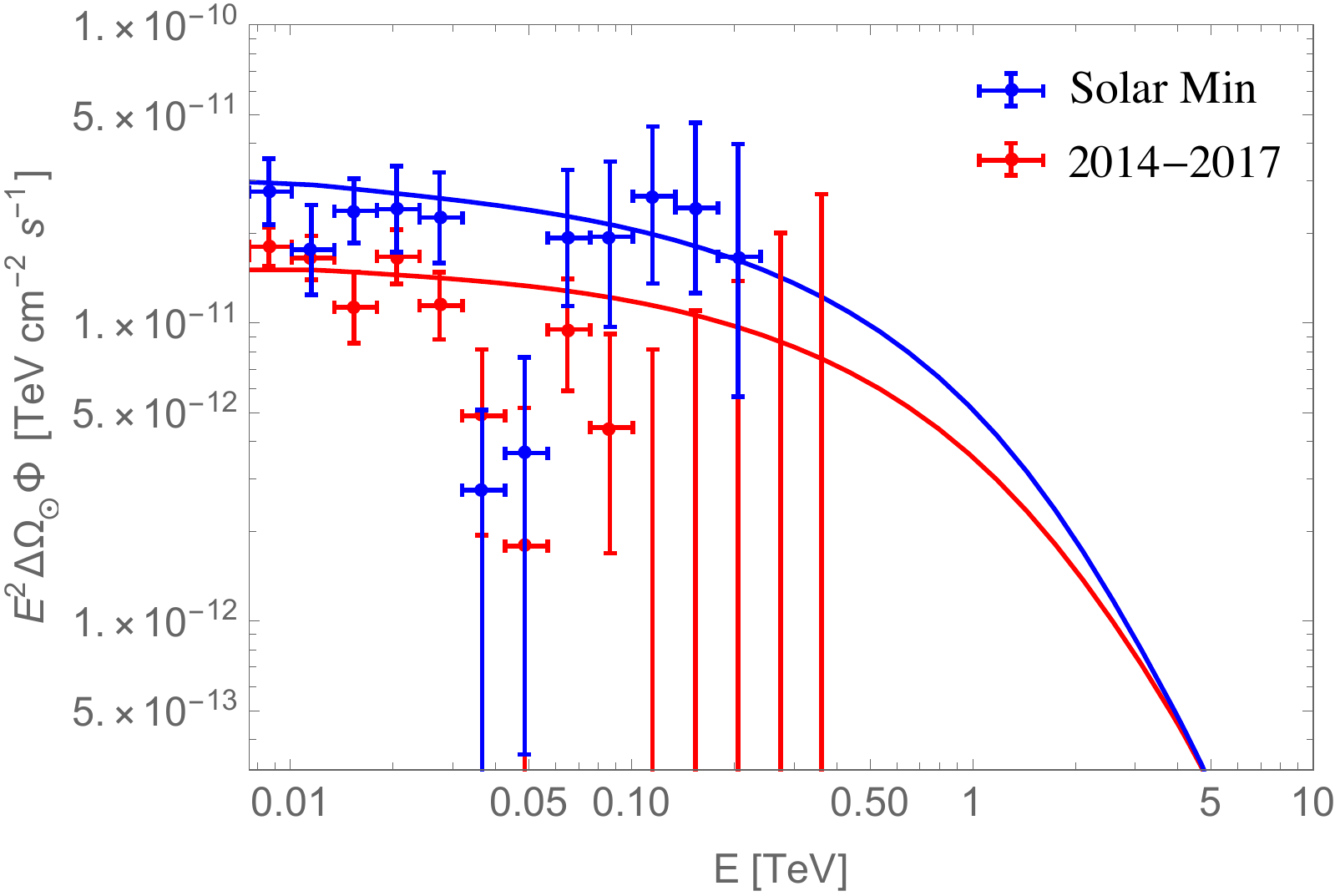}
\end{center}
\vspace{-0.3cm}
\caption{Gamma ray flux from the solar disk  (data from Fermi-LAT  \cite{Linden:2018exo}).
\label{f3}}
\end{figure}

\section{Gamma rays} 
In Fig.~\ref{f3} we plot the flux of gamma rays at $E>10$ GeV implied by our set up
together with the Fermi-LAT data. 
This energy flux exhibits two main features. At low energies it is reduced because primary  CRs do not reach the Sun;
at higher energies it is reduced as well, but because of a different reason: all CRs reach the surface and shower there, but most
photons are emitted towards the Sun. Although the set up does not provide a reason 
for the possible {\it dip} at 40 GeV \cite{Tang:2018wqp}, 
the 400--800 photons per squared meter and year that we obtain seem an acceptable fit of the data.

\begin{figure}[!t]
\begin{center}
\includegraphics[scale=0.556]{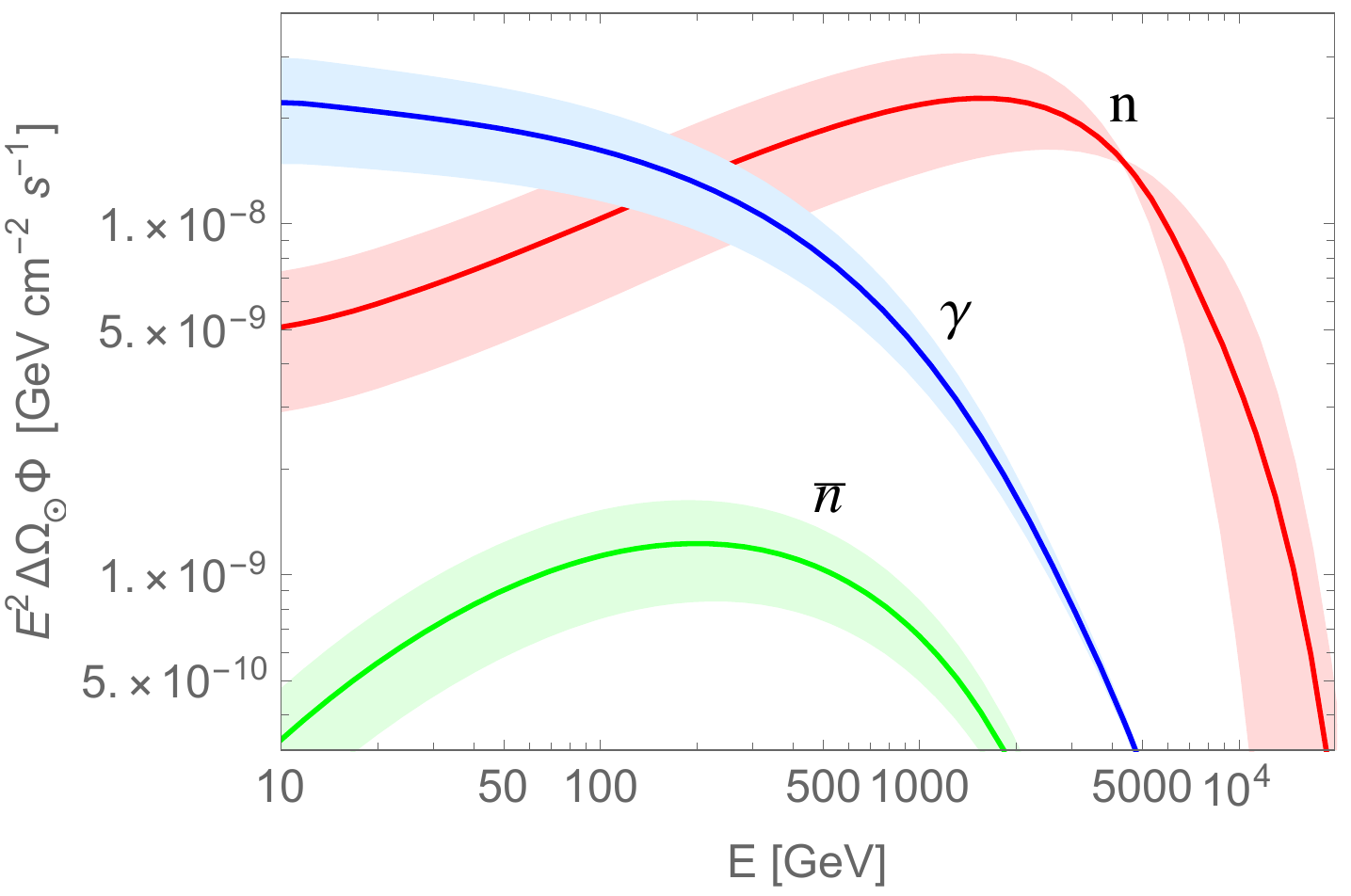} \hspace{0.5cm} \includegraphics[scale=0.56]{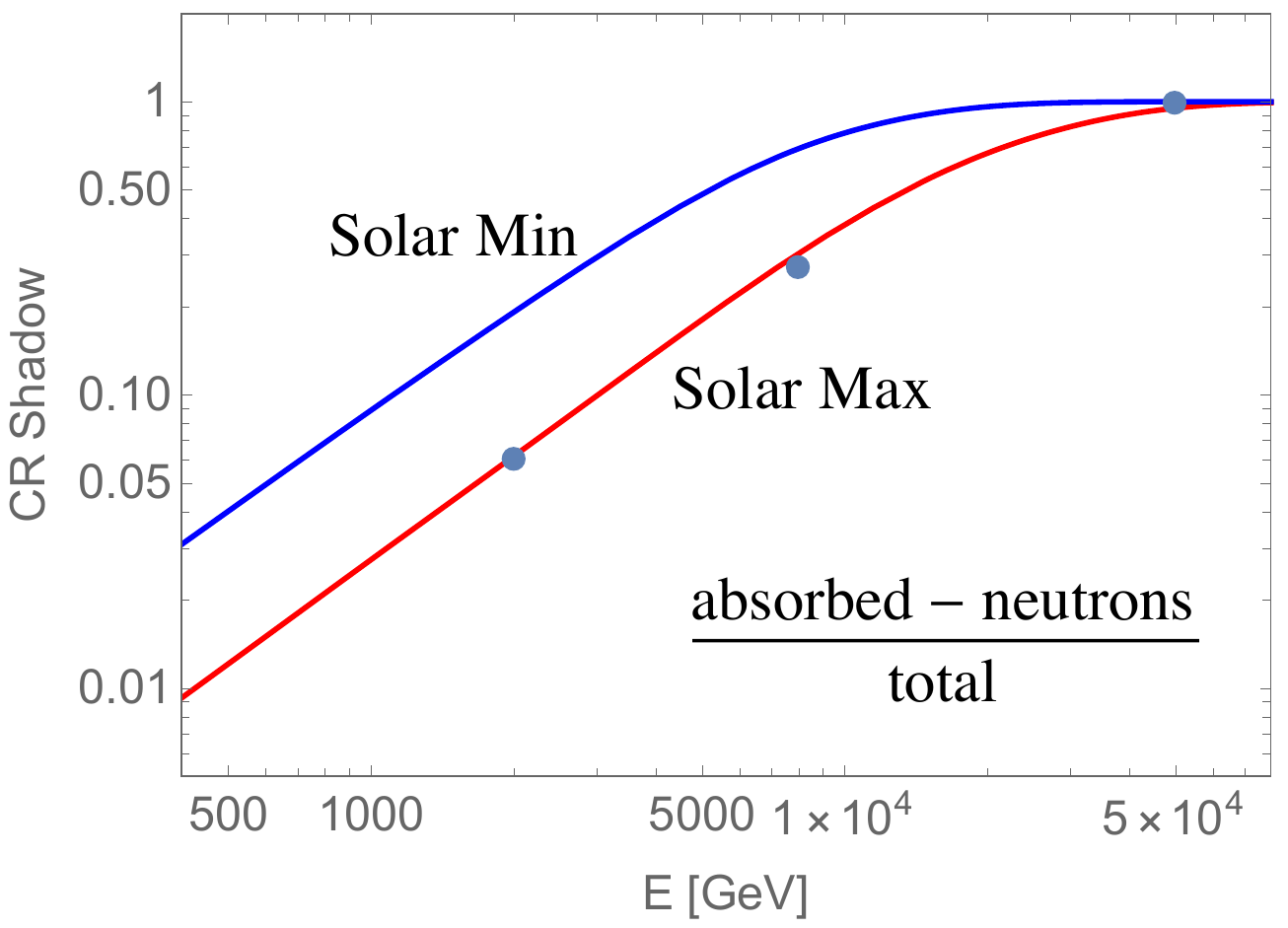}
\end{center}
\vspace{-0.3cm}
\caption{Gamma and neutron fluxes (left) and integrated CR shadow
from the solar disk (right); the dots correspond to the HAWC observation \cite{Enriquez:2015nva}.
\label{f4}}
\end{figure}

\section{Neutrons, CR shadow and muon shadow} 
Our analysis implies an average of 240 neutrons of energy above 10 GeV reaching the Earth from the solar disk
per squared meter and year, with the flux during a solar
minimum a factor of 2 larger than during an active phase of the Sun.
Most of these neutrons come from the spallation of He nuclei (see Fig.~\ref{f4}-left), resulting in a very characteristic 
spectrum that
peaks at 1-5 TeV. The flux is interesting because neutrons are unstable: they can reach us from the Sun, but not from outside
the solar system. In a satellite experiment the background to this solar flux would be the albedo flux from CRs entering the
atmosphere, which seems easily avoidable. Unfortunately, space observatories do not carry hadronic calorimeters and 
are thus unable to detect neutrons.

The solar neutron flux, in turn, has another effect as it enters the atmosphere: it {\it reduces} the CR shadow of the Sun
measured by HAWC. In  Fig.~\ref{f4}-right we give the total shadow (fraction of 
CRs absorbed by the Sun minus the relative number of neutrons reaching 
the Earth) predicted by our framework together with HAWC's data, which was obtained near a solar maximum.

\begin{figure}[t]
\begin{center}
\includegraphics[scale=0.535]{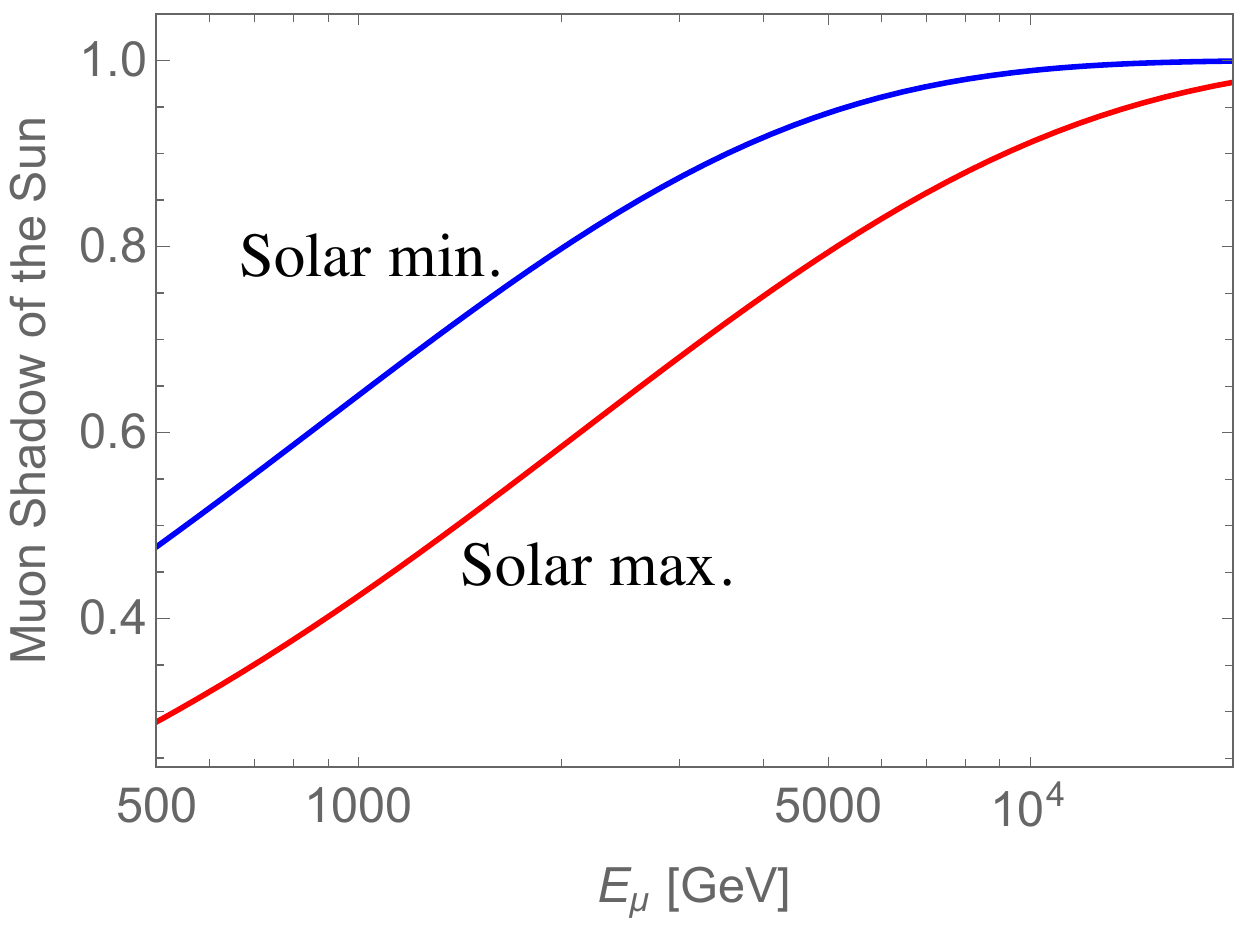} \hspace{0.5cm} \includegraphics[scale=0.55]{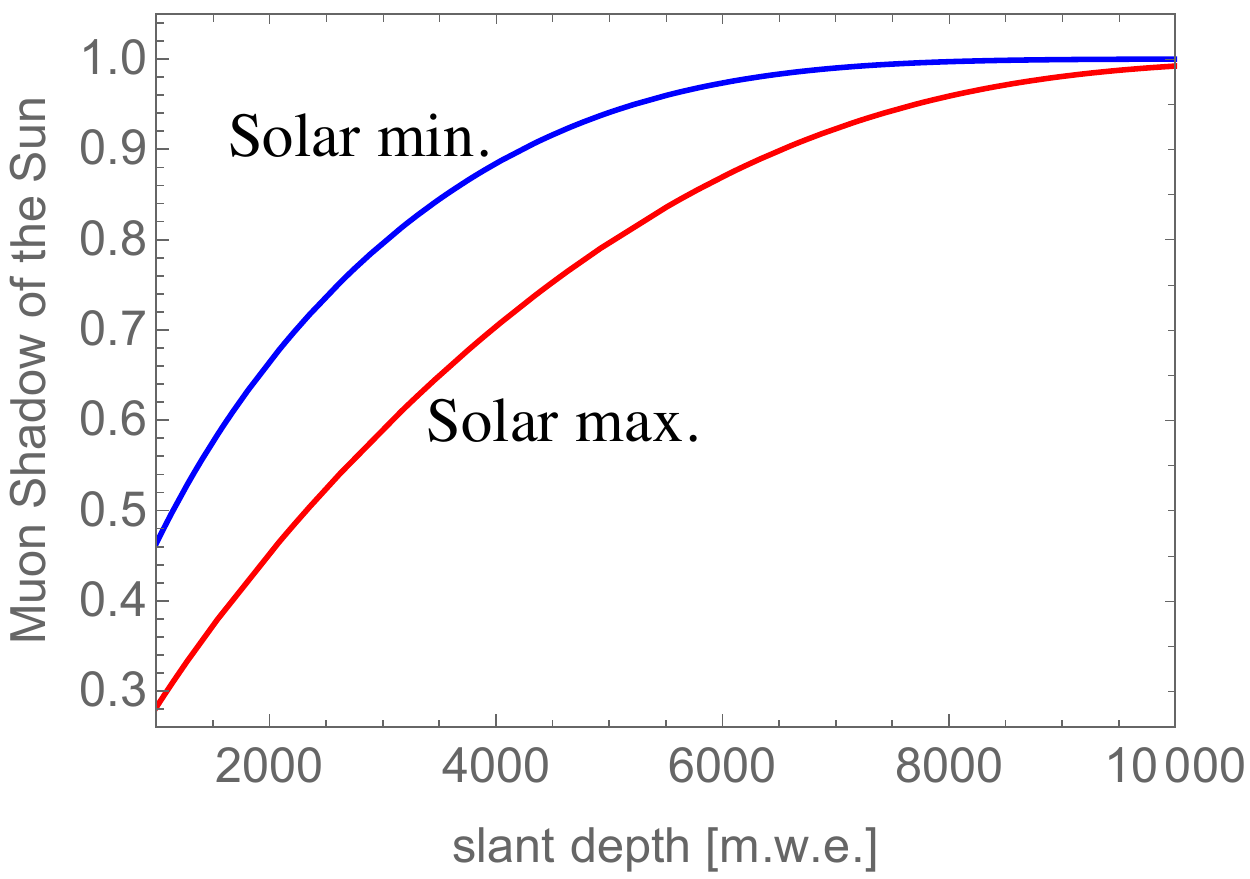}
\end{center}
\vspace{-0.3cm}
\caption{Muon shadow of the Sun at different muon energies (left) or slant depths (right).
\label{f5}}
\end{figure}

In addition to the CR shadow and the gamma and neutron signals, another interesting channel observable 
at neutrino telescopes (already detected at ANTARES \cite{ANTARES:2020yqn} and IceCube \cite{IceCube:2020bsn})
would be the muon shadow of the Sun when it is above the horizon: down-going muons entering
the telescope from the direction of the solar disk. These are muons produced when both the partial shadow of the Sun and the solar
neutrons shower in the atmosphere. In Fig.~\ref{f5} we plot our results as a function of the muon energy (left) or the 
slant depth at the point of entry in the telescope (right).

The plot for the muon shadow at different slant depths is specially revealing. It compares the 
number of tracks from the solar disk (smeared into a larger angular region) and from 
a {\it fake} Sun at the same zenith inclination.
To observe it in a telescope, one should bin the slant depth of the muon tracks entering the detector
and then determine the deficit (integrated to the whole angular region) relative to the fake Sun, finding 
the fraction of a black disk of $r_{\odot}=0.26^\circ$ that it represents. 
In IceCube \cite{Achterberg:2006md} 
the Sun is always very low in the horizon, implying muon tracks of large slant depth and thus always a complete muon
shadow of the Sun. KM3NeT \cite{Adrian-Martinez:2016fdl}, however, 
can access the Sun more vertically and thus from smaller slant depths (down to 3500 m.w.e.), which could establish 
 the energy dependence of this muon shadow.

\section{Neutrinos from the solar disk} 
The neutrino flux reaching a telescope includes three 
differents components: 
\begin{enumerate}
\item Neutrinos produced in the Earth's atmosphere by the partial CR shadow of the Sun. 
At CR energies above 50 TeV the shadow is complete and this component vanishes, but at lower energies the shadow 
disappears and this component should coincide with the atmospheric $\nu$ flux. 
\item Neutrinos produced in the atmosphere by the solar neutrons reaching the Earth.
\item The neutrinos produced in the solar surface, both the albedo flux and the flux from the opposite side that reaches
the Earth after crossing the Sun.
\end{enumerate}

\begin{figure}[!t]
\begin{center}
\includegraphics[scale=0.515]{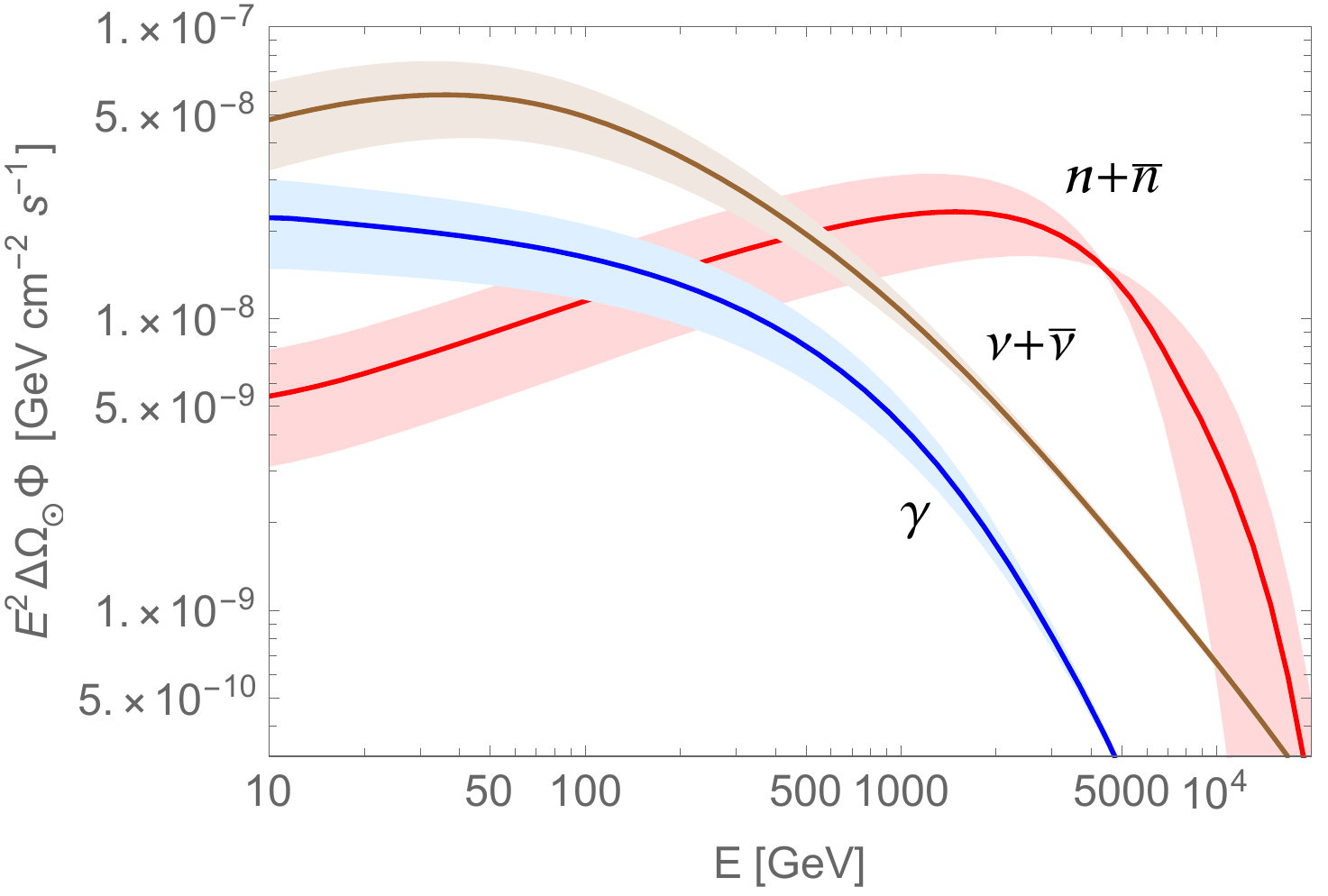} \hspace{0.5cm} \includegraphics[scale=0.52]{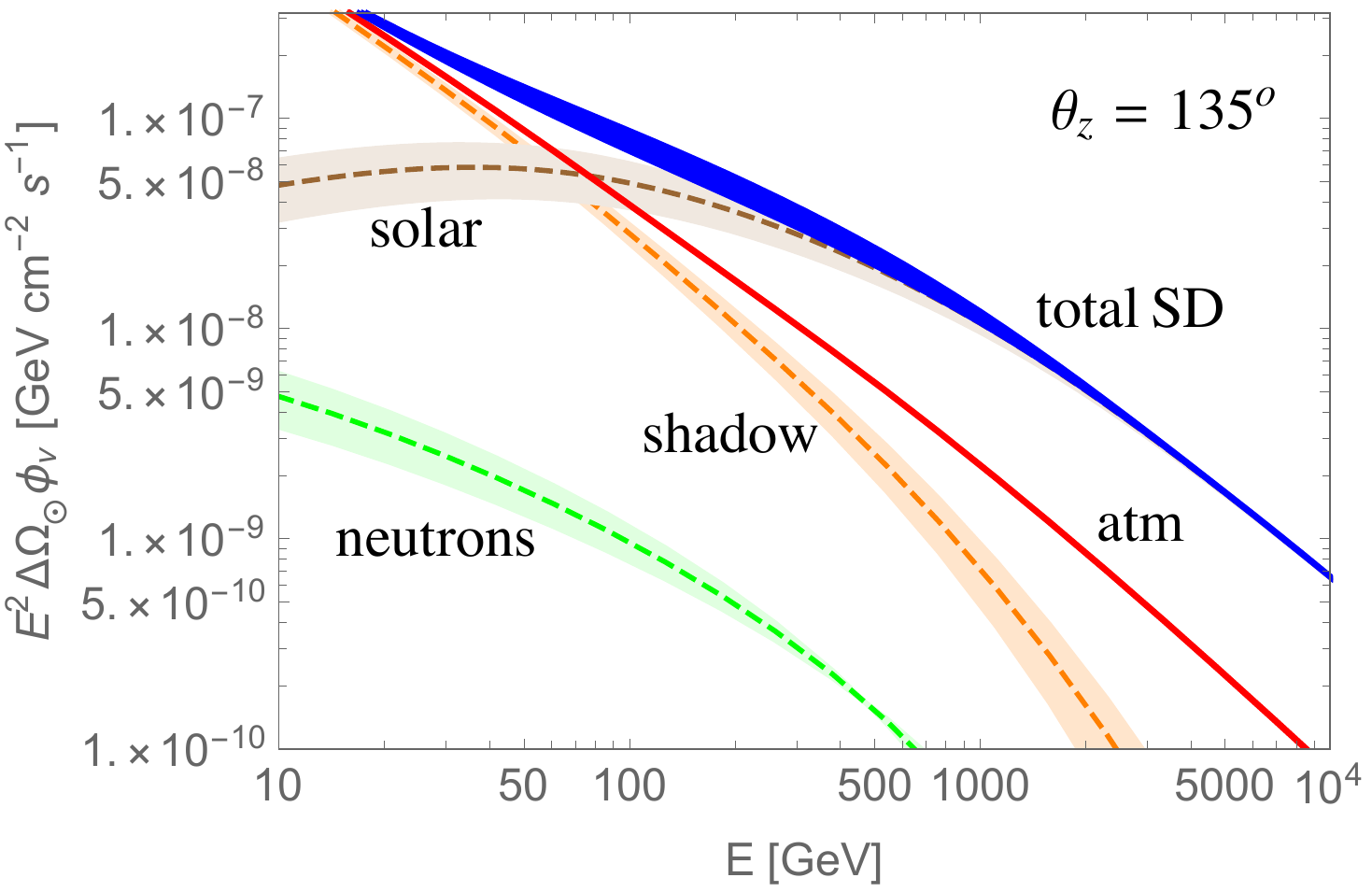}
\end{center}
\vspace{-0.3cm}
\caption{Solar neutrino flux (left) and components defining the neutrino flux 
observed in a telescope from the solar disk at  $\theta_z=135^\circ$ 
and atmospheric background at the same zenith inclination (right).
\label{f6}}
\end{figure}
The first two components are absent in all previous analyses. As for the third one, 
several groups \cite{Edsjo:2017kjk,Ng:2017aur,Arguelles:2017eao} have obtained
the neutrino flux produced by CRs showering in the opposite side of the Sun unaffected by the solar magnetic field 
(see Fig.~\ref{f2}). Their results are larger than ours at energies $E<500$ GeV (in our set up 
low-energy CRs do not reach the Sun), a $30\%$ smaller at 
$E\approx 1$ TeV (our albedo flux is not partially absorbed by the Sun in its way to the Earth) and similar
at $E>10$ TeV (at high energies neutrinos are produced always inwards).
In Fig.~\ref{f6}-left we plot the flux of neutrinos produced in the solar surface together with the albedo flux of gammas and neutrons for comparison. We see that at low energies neutrinos more than 
{\it double} the number of gammas, whereas at $E>5$ TeV all albedo
fluxes vanish but we still get the neutrinos produced in the opposite side of the Sun.

In Fig.~\ref{f6}-right we plot the three neutrino components when the Sun is
$45^\circ$ below the horizon (notice that the fluxes produced by the partial shadow and by the solar neutrons depend
on the zenith angle), together with the atmospheric background. 
The bands express the variation during a solar cicle; the solar and neutron components
are larger during a quiet Sun, whereas the $\nu$ component from the partial shadow is larger during a solar
maximum. The variation in the total neutrino flux during the 11 year cycle (the blue band in the plot) tends to cancel and
is below the 25\% at 200 GeV.

\begin{figure}[!t]
\begin{center}
\includegraphics[scale=0.51]{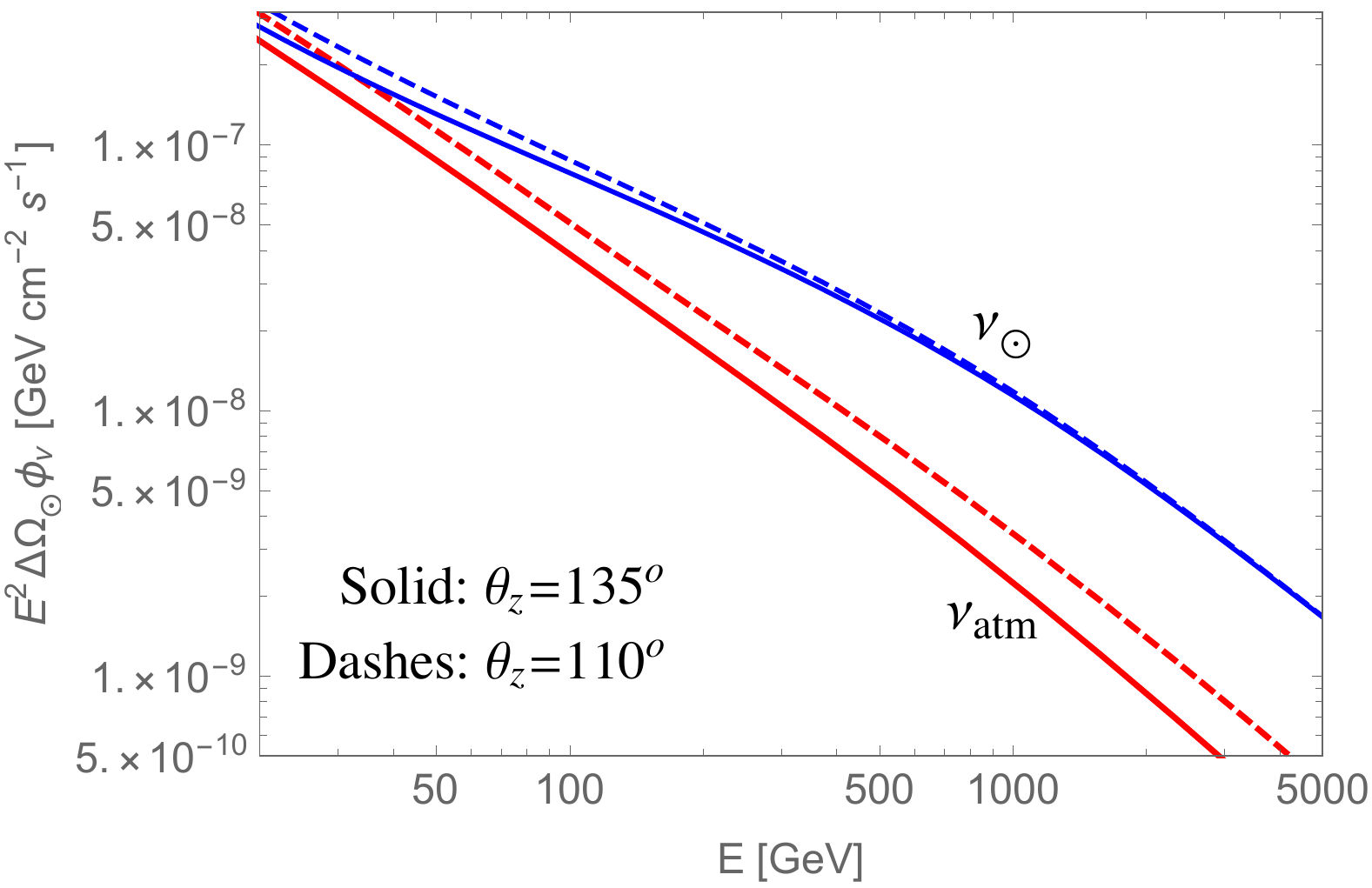}\hspace{0.5cm}\includegraphics[scale=0.50]{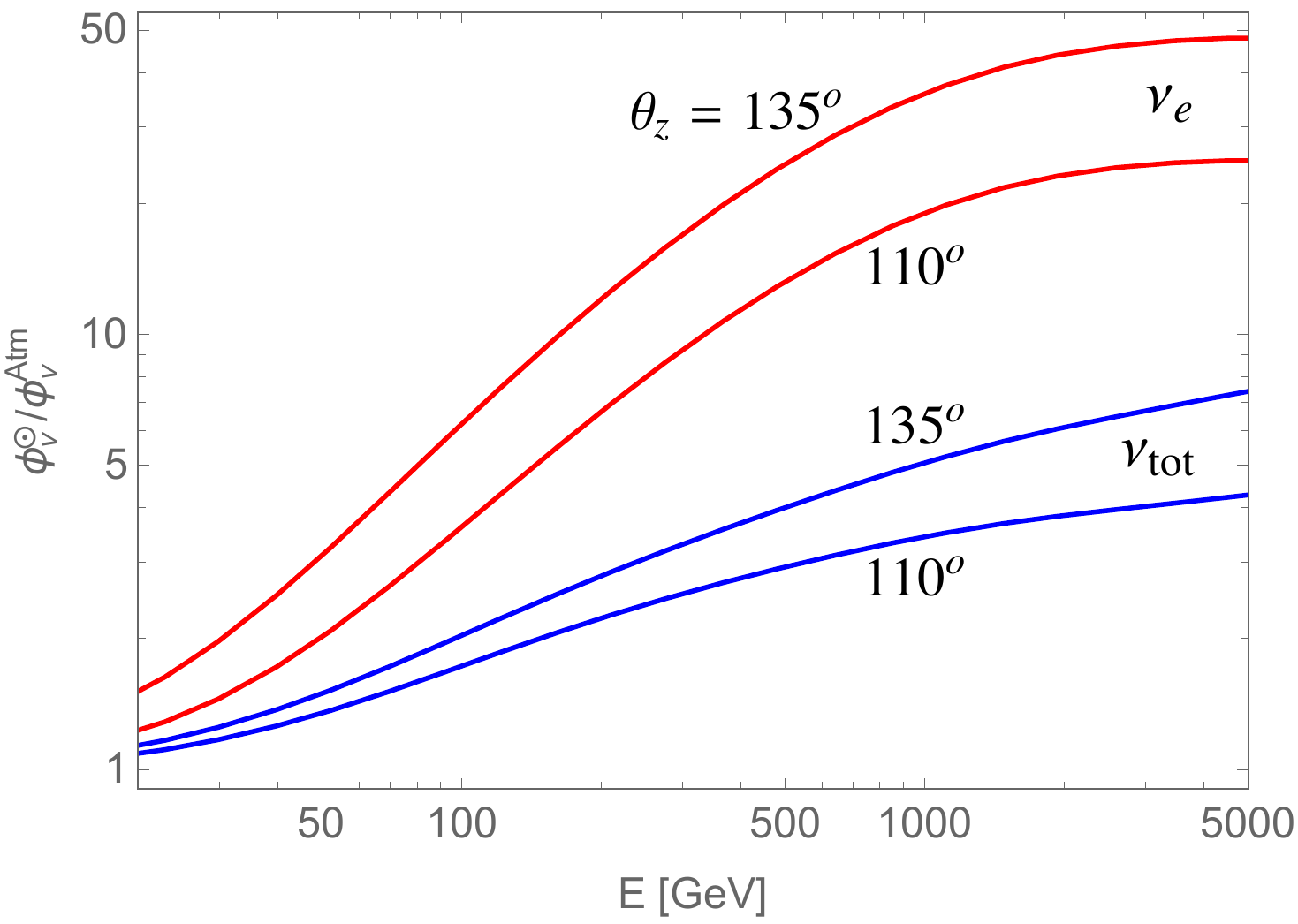}
\end{center}
\vspace{-0.3cm}
\caption{Total neutrino flux from the solar disk  versus atmospheric background for the Sun in
two different zenith inclinations (left).
Signal to background ratio for $\nu_e$ and for the sum of all flavors (right).
\label{f7}}
\end{figure}
We see that the total neutrino flux from the solar disk is well above the atmospheric background at $E>100$ GeV. 
In Fig.~\ref{f7}-left we compare the two fluxes when the Sun is $45^\circ$ or $10^\circ$ below the horizon; the second
inclination is the typical one for the Sun at IceCube. We see that the signal changes little with the zenith angle, whereas the
background is significantly larger when the Sun is near the horizon. In the appendix we provide an analytical fit to these 
components in the neutrino flux, giving the dependence on $\theta_z$ and on the period in the solar cycle.
In Fig.~\ref{f7}-right we plot the signal to background ratio. Since the neutrinos produced in the Sun reach the Earth 
with the same frequency for the three flavors, the ratio is obviously much larger for the $\nu_e$ than the $\nu_\mu$ flavor, and
it grows with the energy and with the zenith inclination.

\section{Solar neutrino floor in indirect searches}  
The annihilation of dark matter (DM) particles $\chi$
captured by the Sun may produce
a $\nu$ flux that, to be detectable, must be above the solar flux just obtained. 
Here we would
like to show how to estimate the minimum DM--nucleon collision cross section that would be accessible in indirect searches.

Let us assume a DM annihilation cross section large enough to establish an stationary regime where the capture rate is equal to
twice the annihilation rate \cite{Jungman:1995df}. 
For illustration, we will consider 3 possible annihilation channels: 
$\chi \bar \chi\to \tau^+ \tau^-$, $\chi \bar \chi\to b \bar b$ and  $\chi \bar \chi\to W^+ W^-$.  
We parametrize the spin-independent (SI) and spin-dependent (SD) DM-nucleon cross sections \cite{Fitzpatrick:2012ix}
\beq
\sigma_{\chi N}^{\rm SI}= {\mu_N^2 \over \pi}\, \left(c_1^N\right)^2\,;\hspace{0.5cm}
\sigma_{\chi N}^{\rm SD}=
{3 \mu_N^2 \over 16 \pi}\, \left(c_4^N\right)^2\,,
\eeq
with $N=p,n$ and $\mu_N=m_N m_\chi/(m_N+m_\chi)$. To deduce the
elastic cross section with the 6 most abundant solar nuclei (H, He, N, O, Ne, Fe) 
we use the nuclear response functions in \cite{Catena:2015uha}. As it is customary in direct searches,
we will take equal proton and neutron SI couplings ($c_1^n=c_1^p$) and will only 
consider the SD coupling of the proton ($c_4^n=0$).

In our estimate of the capture
rate we use the AGSS09 solar model \cite{Asplund:2009fu}
and the SHM$^{++}$  velocity distribution of the galactic DM \cite{Evans:2018bqy}. 
We include the thermal velocity for the solar nuclei, which gives a sizeable contribution ($6\%$ increase) in the
captures through SD interactions (dominated by Hydrogen). We obtain the neutrino yields 
after the propagation from the Sun to the Earth
for each annihilation channel with DarkSUSY \cite{Bringmann:2018lay}.

\begin{figure}[!t]
\begin{center}
\includegraphics[scale=0.48]{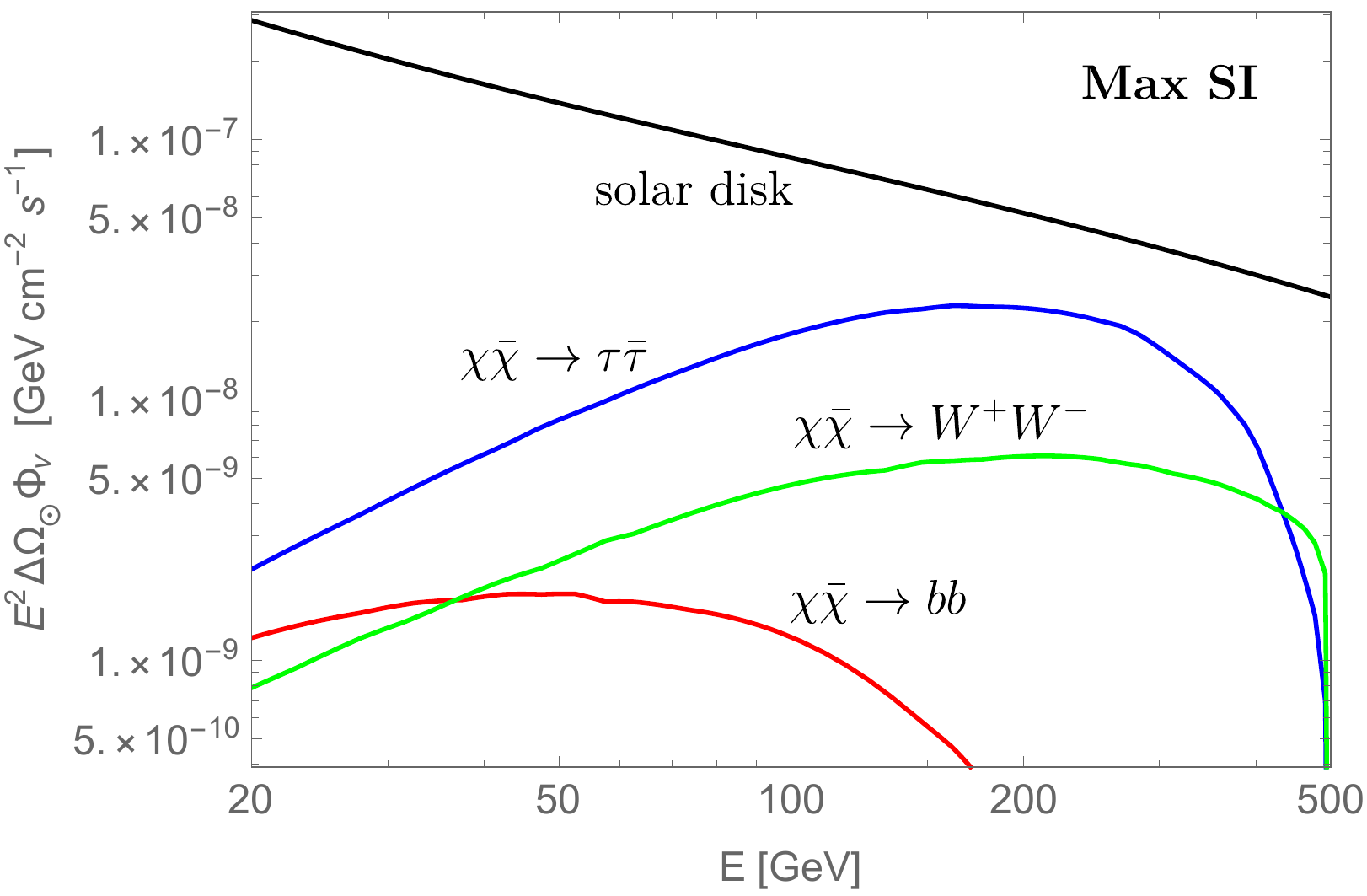}\hspace{0.5cm}
\includegraphics[scale=0.484]{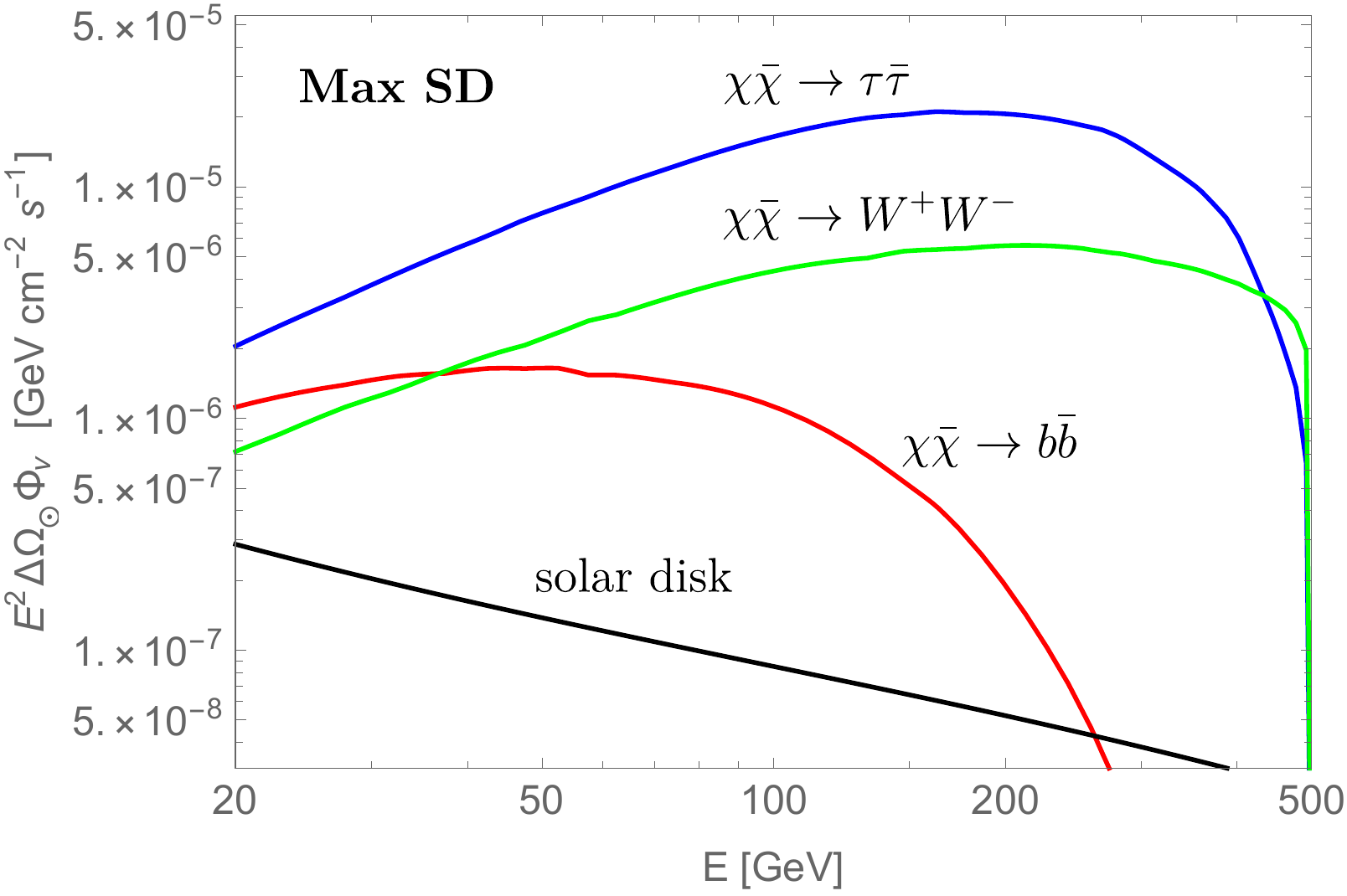}
\end{center}
\vspace{-0.3cm}
\caption{Maximum neutrino flux from DM annihilation consistent with the bounds on $\sigma_{\chi N}^{\rm SI}$ (left)
and $\sigma_{\chi N}^{\rm SD}$ (right) from
direct searches for the three annihilation channels that we have considered.
\label{f9}}
\end{figure}
To illustrate the reach of DM searches at $\nu$ telescopes, let us fix $m_\chi=500$ GeV.
For a DM particle that is captured through SI collisions and 
annihilates into $\tau^+ \tau^-$, a neutrino flux above the solar background established
in the previous section requires $\sigma_{\chi N}^{\rm SI}>9.7\times 10^{-46}$ cm$^2$. 
If the annihilation channels are $W^+ W^-$ and $b \bar b$ then the elastic cross section must
be $\sigma_{\chi N}^{\rm SI}>3.1\times 10^{-45}$ cm$^2$ 
and $\sigma_{\chi N}^{\rm SI}>3.0\times 10^{-44}$ cm$^2$, respectively. These cross sections,
however, are already excluded by direct searches at XENON1T \cite{XENON:2018voc}:  
$\sigma_{\chi N}^{\rm SI}<4.4\times 10^{-46}$ cm$^2$. 
In Fig.~\ref{f9} we plot the neutrino fluxes at the Earth for this maximum value of 
$\sigma_{\chi N}^{\rm SI}$ together with the solar background. The fluxes from
DM annihilation are below the solar background in the whole energy range. If DM had only SI interactions with matter, and 
it had this mass and annihilation channels, indirect searches would reach the solar neutrino floor before
they discover it. 

Indirect searches could discover DM only if it had monochromatic annihilation channels ({\it e.g.,} 
$\chi\bar \chi \to \nu X$)  or a large SD cross section, which is much less constrained. In particular, PICO-60 establishes that
$\sigma_{\chi p}^{\rm SD}<1.8\times 10^{-40}$ cm$^2$ at $m_\chi=500$ GeV \cite{PICO:2019vsc}, while a DM that annihilates into
$W^+W^-$ requires just $\sigma_{\chi p}^{\rm SD}>1.6\times 10^{-42}$ cm$^2$ to be above the flux of solar neutrinos.

\section{Summary and discussion}  
TeV CRs induce an indirect solar emission that was discussed more than 40 years ago and has already been detected in gamma rays. Here we have used the energy dependence of the shadow of the Sun at HAWC to define a set up that 
implies very definite fluxes of gammas, neutrons and neutrinos. The set up explains the peculiar 
spectrum of the solar gammas observed at Fermi-LAT's: low energy CRs do not contribute to the albedo flux as they do not 
reach the solar surface, whereas high energies CRs reach the Sun, but they produce gammas that are emitted 
mostly inwards and never reach the Earth.
As we discuss in Section 4, our framework could be confirmed if KM3NeT observed a slant-depth dependence in the muon shadow of the Sun. We propose up to five different (but related) signals whose observation would draw a more complete picture of the 
solar magnetism and of the propagation of TeV CRs near the surface.

The neutrino fluxes from the solar disk are specially interesting, as there is currently an important experimental effort for indirect DM searches.
We show that the neutrino fluxes reaching a telescope include three components: {\it (i)} the solar emission from both sides of the Sun, {\it (ii)} neutrinos produced when the partial CR shadow of the Sun enters the atmosphere, and {\it (iii)} neutrinos produced also in the atmosphere by the albedo flux of solar neutrons. The two last contributions have not been discussed in previous literature.
In the appendix we provide fits for these components, giving their explicit dependence on the zenith angle and the period of the 11 year solar cycle.

These neutrinos 
define a floor in DM searches. In particular, we find that the maximum SI elastic cross section consistent with current bounds from XENON1T implies a flux of neutrinos from DM annihilations into $\tau^+ \tau^-$, $b \bar b$ or $W^+ W^-$  
already below this floor. Therefore, 
a precise characterization of the $\nu$ fluxes from the solar disk induced by CRs
is essential both to decide the optimal detection strategy  and to establish the reach of indirect DM searches at each neutrino telescope.

\section*{Acknowledgments}
This work was partially supported by the Spanish Ministry of Science, Innovation and Universities
(PID2019-107844GB-C21/AEI/10.13039/501100011033) and by the Junta de Andaluc{\'\i}a 
(FQM 101, P18-FR-5057).

\appendix

\section{Fit to the neutrino fluxes}  
Here we provide approximate fits for the atmospheric and solar fluxes integrated over the angular region 
($\Delta\Omega_\odot$, with a radius of $0.26^\circ$) occupied by the Sun. In these expressions $E$ is in GeV and $t$ in years
($t=0$ at the solar minimum), whereas $\Delta\Omega_\odot \, \Phi_{\nu_\mu}^{\rm atm}$ is given in GeV$^{-1}$ cm$^{-2}$ s$^{-1}$.
The angle $\theta^*(\theta_z)$ is defined in \cite{Lipari:1993hd,Gutierrez:2021wfy}:
\beq
\tan \theta^* = {R_\oplus \sin\theta_z\over \sqrt{R_\oplus^2 \cos^2\theta_z + \left( 2 R_\oplus + h \right) h }}\,.
\label{theta}
\eeq
For the atmospheric flux we have
\beq
\Delta\Omega_\odot \, \Phi_{\nu_\mu}^{\rm atm} (E,\theta)= 
1.05\times 10^{3} \;E^{-2.97 - 0.0108 \log E - 
  0.00141 \log^2 E} \,F^{\rm atm}_1(E,\theta)\,;
\eeq
\beq
\Delta\Omega_\odot \, \Phi_{\nu_e}^{\rm atm} (E,\theta)= 
460 \;E^{-3.30 - 0.0364 \log^{1.35} E +
0.0103 \log^{1.85} E}\,F^{\rm atm}_2(E,\theta)\,,
\eeq
with
\beq
F^{\rm atm}_1(E,\theta)={\left( {176\over E} \right)^{0.6} + 
\cos [\theta^*({\pi\over 4})] \over \left( {176\over E} \right)^{0.6}  + \cos[\theta^*(\theta_z)]}\,;\;\;
F^{\rm atm}_2(E,\theta)={\left( {7.5\times 10^{-4}\over E} \right)^{0.21} + 
\cos [\theta^*({\pi\over 4})] \over \left( {7.5\times 10^{-4}\over E} \right)^{0.21}  + \cos[\theta^*(\theta_z)]}\,.
\eeq
For the atmospheric neutrinos from both the CR shadow of the Sun and solar neutrons,
\beq
\Delta\Omega_\odot \, \Phi_{\nu_\mu}^{\rm shad+n}(E,\theta,t) = 
1.04\times 10^{3} \;E^{G^{\rm atm}_1(E,t)} \,F^{\rm atm}_1(E,\theta)\,;
\eeq
\beq
\Delta\Omega_\odot \, \Phi_{\nu_e}^{\rm shad+n}(E,\theta,t) = 
327 \;E^{G^{\rm shad+n}_2(E,t)}\,F^{\rm atm}_2(E,\theta)
\eeq
with
\beq
G^{\rm shad+n}_1(E,t)=-2.98 - 0.017 \log E +
 0.012 \cos {2\pi t\over 11} \log^2 E -3.3\times 10^{-4} \log^3 E-
 4.1\times 10^{-6} \log^5 E\,;
\eeq
\beq
G^{\rm shad+n}_2(E,t)=-3.1 - 0.061 \log E - \cos {2\pi t\over 11} \left(
0.00305 \log E  +2.1\times 10^{-6} \log^5 E\right) - 5.3\times 10^{-7} \log^6 E\,.
\eeq
Finally, the neutrinos produced in the Sun come in the three flavors with the same frequency and
\beq
\Delta\Omega_\odot \, \Phi_{\nu_i}^{\rm \odot} (E,t)= 
\left( 1.50 - {572\sin^2 {\pi t\over 11}\over 900+E}  \right)
E^{-1.20 - 0.1 \log E - 
  0.0042 \log^2 E+ 1.6\times 10^{-5} \log^4 E} \,.
\eeq

\end{document}